\documentclass[twocolumn]{aastex7}

\newcommand{\msun}{\ensuremath{\textrm{M}_{\odot}}}
\newcommand{\mstar}{\ensuremath{\textrm{M}_{\ast}}}
\newcommand{\lgmstar}{\ensuremath{\log_{10}(\mstar/\msun)}}

\newcommand{\dbalmer}{\ensuremath{\textrm{D}_{n}(4000)}}

\newcommand{\ewha}{\ensuremath{\textrm{EW}(\textrm{H}\alpha)}}

\usepackage{soul}

\usepackage{subfigure}

\shorttitle{Star Formation Cessation Criteria}
\shortauthors{Cheng et al.}

\begin{document}

\title{SDSS-IV MaNGA: Star Formation Cessation in Low-redshift Galaxies. III. Dependence on Quenching Criteria}

\author[orcid=0009-0001-5437-410X]{Zhuo Cheng}
\affiliation{Department of Physics, The Chinese University of Hong Kong, Sha Tin, NT, Hong Kong, China}
\email[show]{chengzphy@gmail.com}

\author[orcid=0009-0004-6271-4321]{Tao Jing}
\affiliation{Department of Astronomy, Tsinghua University, Beijing 100084, China}
\email{jingt20@mails.tsinghua.edu.cn}

\author[orcid=0000-0002-8711-8970]{Cheng Li}
\affiliation{Department of Astronomy, Tsinghua University, Beijing 100084, China}
\email[show]{cli2015@tsinghua.edu.cn}    

\author[orcid=0000-0003-1025-1711]{Renbin Yan}
\affiliation{Department of Physics, The Chinese University of Hong Kong, Sha Tin, NT, Hong Kong, China}
\affiliation{CUHK Shenzhen Research institute, No.10, 2nd Yuexing Road, Nanshan, Shenzhen, China}
\email{rbyan@cuhk.edu.hk} 

\correspondingauthor{Zhuo Cheng, Cheng Li}
\begin{abstract}

This paper is the third in a series of studies investigating star formation cessation in nearby galaxies on kiloparsec scales. Using the final SDSS-IV MaNGA data release, we ask how the inferred importance of global, local, and environmental properties depends on the operational definition of quenched regions. We classify spaxels as star-forming, reliably quenched, or potentially quenched by accounting for measurement uncertainties, and train random forest classifiers with a parameter set chosen for direct comparison with previous work. For reliably quenched regions, the local stellar mass surface density $\Sigma_\ast$ consistently has the highest feature importance, independent of quenching definition. By contrast, the high importance of central velocity dispersion $\sigma_c$, previously interpreted as evidence for galaxy-wide AGN feedback, is recovered mainly when potentially quenched regions are included. The leading parameter also varies with stellar mass: $\Sigma_{\rm 1kpc}$ is most important below $\sim10^{10.2}\,\msun$, whereas local quantities such as $\Sigma_\ast$ and $\sigma_\ast$ become more prominent at high masses. These results show that quenching criteria and uncertainty treatment can reconcile apparently discrepant feature-importance studies. AGN-related processes may contribute to ambiguous regions, but the reliably quenched population is most tightly linked to high local stellar density.

\end{abstract}

\keywords{Galaxy evolution --- Galaxy quenching --- Star formation --- Integral field spectroscopy}


\section{Introduction} \label{sec:intro}
The cessation of star formation is a central problem in galaxy evolution, as the fraction of red quiescent galaxies has increased substantially since $z\sim1$ \citep[e.g.][]{Bell2004, Bundy2006, Faber2007, Muzzin2012}. Star formation can cease when cold gas is removed, heated, or prevented from collapsing efficiently, but the relative importance of these pathways remains debated. Proposed mechanisms include halo shock heating, ram-pressure stripping, tidal effects, strangulation, and mergers \citep[e.g.][]{Rees1977, Birnboim2003, Dekel2006, Gunn1972, Moore1996, Hopkins2006}, as well as internal processes such as stellar feedback, AGN feedback, dynamical stabilization, morphological quenching, and angular-momentum related suppression \citep[e.g.][]{Cox1974, Fabian2012, Toomre1964, Martig2009, Obreschkow2016}. These mechanisms operate on different spatial scales and may leave different signatures in resolved galaxy data.

Spatially resolved spectroscopy is therefore essential for connecting star formation cessation to physical conditions within galaxies. Integral field unit (IFU) surveys provide measurements of stellar populations, emission lines, and kinematics on kiloparsec scales \citep[e.g.][]{de2002, Cappellari2011, sa2012, Bryant2015, Bundy2015}, allowing quenching to be studied locally rather than only through central or galaxy-integrated measurements. For instance, studies of MaNGA galaxies have found evidence for inside-out star formation cessation in relatively massive galaxies \citep[e.g.][]{Li2015,Wang2018}, as well as centrally suppressed and more extended patterns of star formation suppression in nearby galaxies \citep[e.g.][]{Belfiore2017, Ellison2018, Spindler2018}. Resolved spectroscopy has also shown that kiloparsec-scale star formation and low-ionization emission are closely linked to local stellar mass density \citep{Hsieh2017}.

A central complication, however, is that a ``quenched'' region is not uniquely defined in resolved data. H$\alpha$ emission traces ongoing star formation but can be affected by non-star-forming ionization, while \dbalmer, the narrow-definition 4000\AA\ break \citep{Balogh1999}, is sensitive to young stellar populations formed within the past 1--2~Gyr \citep{Kauffmann2003}. Quenched regions can also be selected through low-ionization emission-line classifications in the BPT diagram \citep{Baldwin1981, Belfiore2017}. For example, \citet{Li2015} and \citet[][hereafter Paper~I]{Wang2018} defined quenched regions using $\ewha<2$\AA\ and $\dbalmer>1.6$, whereas \citet{Lin2019b} selected quenched regions using LI(N)ER classification on the BPT diagram together with weak H$\alpha$ emission ($\ewha<3$\,\AA). \citet{Bluck2020a, Bluck2020b} instead identified quenched spaxels by their offset below the empirical resolved star-forming main sequence; in practice, their non-SF selection corresponds to broad-definition $D(4000)>1.45$, which corresponds approximately to narrow-definition $\dbalmer>1.68$ using the scaling of \citet{Gorgas1999}. 

Since different diagnostics carry different assumptions about timescale, ionizing source, and required emission-line S/N, quenched regions defined by different diagnostics may lead to discrepant conclusions. The random forest analysis of quenched regions in \citet{Bluck2020a, Bluck2020b} found central velocity dispersion to have the highest feature importance, which was interpreted as evidence for galaxy-wide AGN feedback; related MaNGA kinematic work reached a similar conclusion for velocity dispersion at the galaxy scale \citep{Brownson2022}. In contrast, \citet[][hereafter Paper~II]{Jing2024} found that local stellar mass surface density ($\Sigma_\ast$) was most important for quenched yet gas-rich regions in disk galaxies, suggesting a close connection with local stellar density and evolved-stellar-population-related processes. A recent MaNGA DR17 comparison found that the inferred inside-out and outside-in quenching fractions depend strongly on whether suppression is traced by sSFR, \dbalmer, post-starburst features, or LI(N)ER emission \citep{Ho2026}. 

A second, closely related issue is how to handle weak-line spaxels. BPT-based classifications require detections in several emission lines, whereas old or partially quenched regions often have weak H$\alpha$ emission and uncertain line ratios. Treating all such spaxels as ordinary quenched regions can therefore mix robustly quenched regions with ambiguous low-S/N measurements.

This paper tests whether the discrepant results above can be traced to different quenching definitions and to the different treatments of ambiguous low-S/N spaxels. We adopt two literature-based quenching definitions, corresponding to the Paper~I/Paper~II and \citet{Bluck2020a} selections. We do not consider the \citet{Lin2019b} definition separately because Paper~II showed that its LI(N)ER+\ewha\ selection gives results similar to the \dbalmer+\ewha\ definition used in Papers~I and II. We separate reliably quenched regions from potentially quenched regions whose classification is sensitive to measurement uncertainties, and train random forest classifiers using a feature set chosen to match \citet{Bluck2020a}. We then quantify how feature importance changes between reliably and potentially quenched samples and across stellar mass and galactocentric radius. Throughout the paper, we adopt a $\Lambda$CDM cosmology with $\Omega_m=0.3$ and $\Omega_\Lambda=0.7$, and a Hubble constant of $H_0=70~{\rm km~s^{-1}~Mpc^{-1}}$.

\section{Data} \label{sec:data}
\subsection{MaNGA} \label{subsec:manga}

MaNGA is one of the three major programs of SDSS-IV \citep{Blanton2017} and obtained integral field spectroscopy for 10,010 nearby galaxies \citep{Bundy2015}. Targets were selected from the NASA Sloan Atlas \citep[NSA;][]{Blanton2011} with $0.01<z<0.15$ and $5\times10^8\msun\leq \mstar\leq 3\times10^{11}\msun$ \citep{Wake2017}. The survey used 17 fiber-bundle IFUs with fields of view from $12^{\prime\prime}$ to $32^{\prime\prime}$, producing datacubes with $0.5^{\prime\prime}$ spaxels and an effective spatial resolution of $\sim2.5^{\prime\prime}$ \citep{Drory2015,Law2015}. The spectra cover 3622--10354\AA\ at $R\sim2000$ \citep{Smee2013}.

The MaNGA datacubes are produced by the Data Reduction Pipeline \citep{Law2016}; stellar kinematics, emission-line measurements, and spectral indices are derived by the Data Analysis Pipeline \citep{Westfall2019}. Details of the flux calibration, observing strategy, and data quality are given by \citet{Yan2016a,Yan2016b}. We use the final SDSS-IV data release \citep{SDSS_DR17}.

\begin{figure*}[!t]
\centering
    \includegraphics[width=0.94\linewidth]{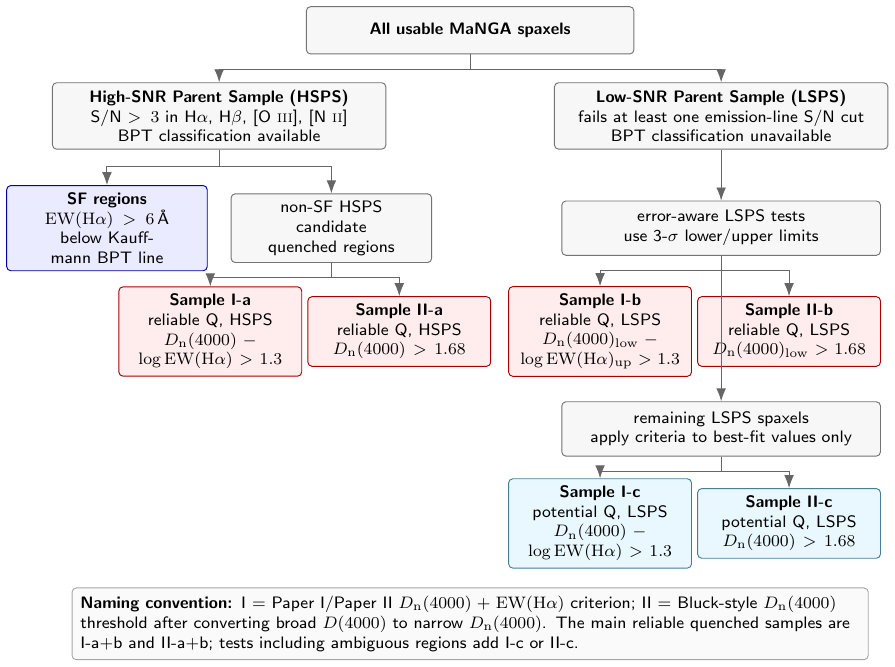}
\caption{Flowchart summarizing the spaxel classification and sample naming. The first split separates the High-SNR Parent Sample (HSPS), where BPT classification is available, from the Low-SNR Parent Sample (LSPS), where at least one required emission line lacks S/N $>3$. In HSPS, SF regions are selected with the BPT diagram and H$\alpha$ equivalent width, while non-SF spaxels can enter the reliably quenched samples. In LSPS, reliably quenched samples require the relevant criterion to remain satisfied after applying 3-$\sigma$ measurement uncertainties, whereas potentially quenched samples satisfy the same criteria only for the best-fit measurements after reliably quenched regions are removed.}
\label{fig:sample_flowchart}
\end{figure*}

\subsection{Regional parameters from MaNGA spaxels}  \label{subsec:data_pre}

For this work, we use spaxel-level stellar population, emission-line, kinematic, and dust measurements from \citet{Li-Li-2023}. Their analysis fits each MaNGA spectrum with PCA-compressed stellar templates based on the \citet[][BC03]{Bruzual2003} models, subtracts the stellar continuum, and measures emission-line fluxes and equivalent widths from Gaussian fits to the residual spectrum. Dust attenuation is corrected using the Balmer decrement for emission lines and the method of \citet{Li2020} for stellar continua. We use the resulting \dbalmer, \ewha, and emission-line fluxes to define the spaxel samples and construct diagnostic diagrams.

We use the following four regional properties for the random forest analyses: $\Sigma_\ast$ (stellar surface mass density, corrected for projection effects), $Z_\ast$ (light-weighted stellar metallicity), $\sigma_\ast$ (stellar velocity dispersion), and $R/R_e$ (galactocentric distance normalized by the effective radius).

\subsection{Host galaxy and environment parameters} \label{subsec:host_env}

Host-galaxy and environmental properties are taken from NSA and MaNGA Value Added Catalogs. These include stellar masses from NSA, morphological T-types from \citet{Dom2018}, bulge-to-total luminosity ratios from \citet{Meert2015}, and MaNGA-GEMA environmental measurements\footnote{\url{https://data.sdss.org/datamodel/files/MANGA_GEMA/GEMA_VER/GEMA.html}}: group halo mass ($M_{h}$), local density $\log(1+\delta)$, and satellite distance $r/r_{200}$ \citep{Yang2007, Wang2009, Wang2012}.

The global input properties used in the random forest analyses are $M_\ast$ (total stellar mass), $B/T$ (bulge-to-total luminosity ratio), $\sigma_c$ (stellar velocity dispersion at the galactic center), and $\Sigma_{\rm 1kpc}$ (stellar surface mass density within the central 1~kpc). We estimate $\Sigma_{\rm 1kpc}$ using SDSS imaging data rather than MaNGA data, as the relatively poor spatial resolution of MaNGA leads to underestimated $\Sigma_{\rm 1kpc}$ estimates at $z\ga 0.03$ (see~\autoref{sec:app}). For each galaxy, we derive $\Sigma_{\rm 1kpc}$ by applying the $M_\ast/L$ versus $g-r$ relation of \citet{Du2019}, calibrated from a joint analysis of SDSS imaging and MaNGA IFS data. 

The environmental input properties are $M_h$ (dark matter halo mass of the SDSS galaxy group that hosts the galaxy), $\log(1+\delta)$ (average local density within 1~Mpc), and $r/r_{200}$ (distance of satellite galaxies to the central galaxy of their host halo, normalized by the halo radius; this parameter is set to zero for central galaxies).

\begin{figure*}[ht]
\centering
	\subfigure{\label{fig:subfig:a1}}
	\includegraphics[width=0.4\linewidth]{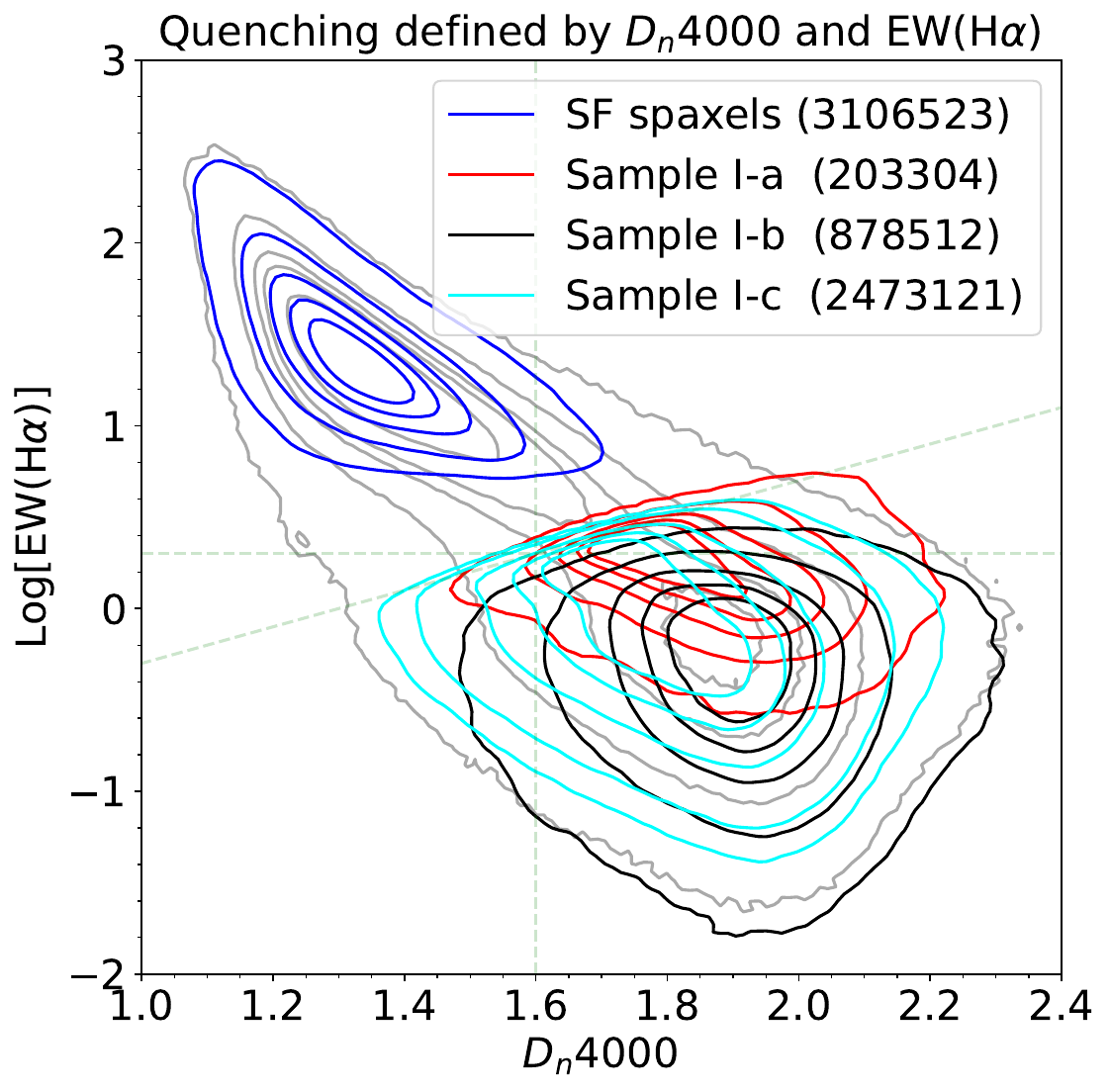}
	\hspace{0.01\linewidth}
	\subfigure{\label{fig:subfig:a2}}
	\includegraphics[width=0.4\linewidth]{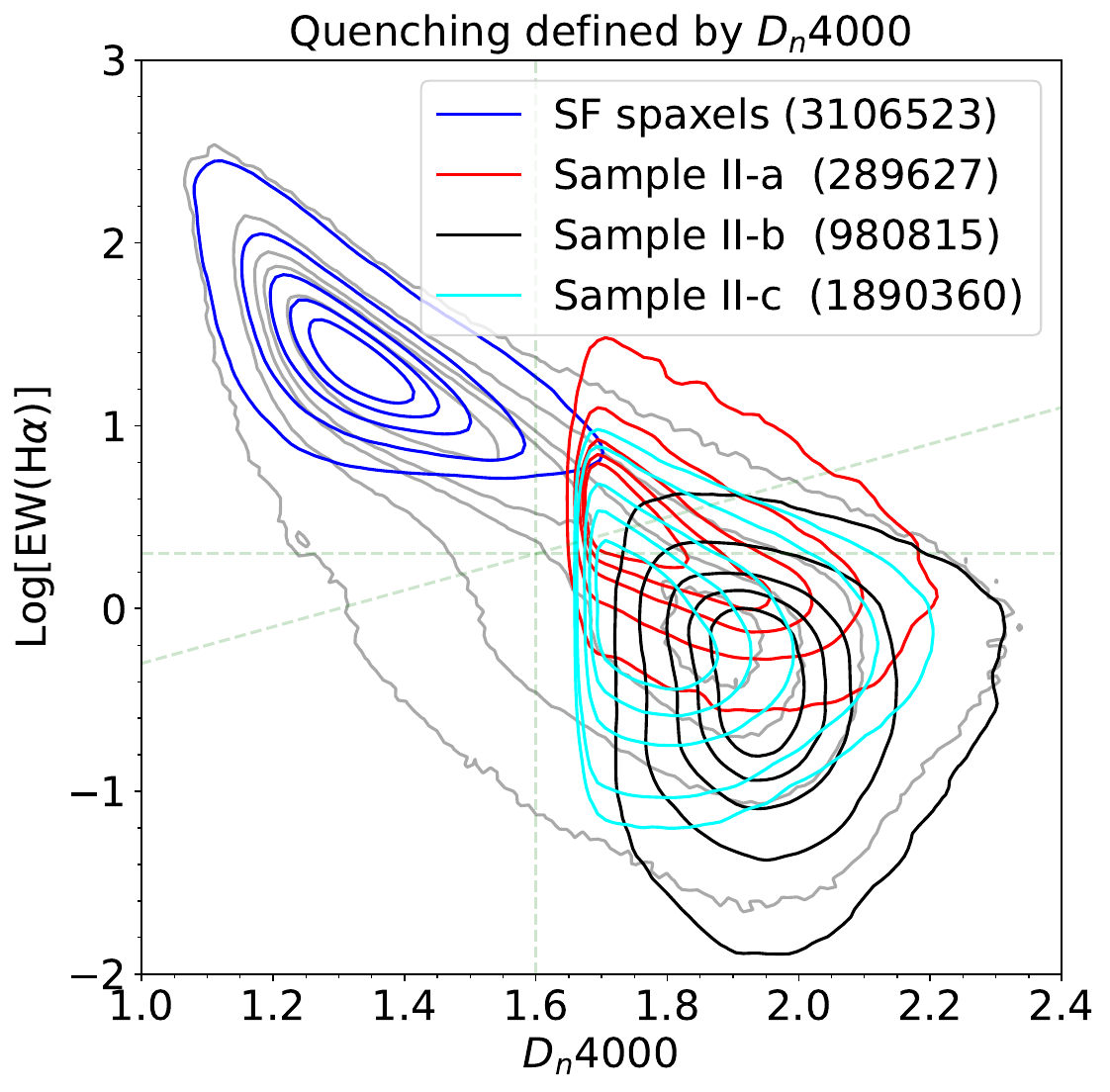}
\caption{Contour distributions of the spaxel subsamples in the plane of $\log_{10}$[EW(H$\alpha$)] versus \dbalmer. Gray contours show the parent sample, while colored contours correspond to the subsamples listed in each legend. Numbers in parentheses indicate the number of spaxels in each subsample.}
\label{fig:d4000_ha_plane}
\end{figure*}

\begin{figure*}[!tp]
\centering
    \subfigure{\label{fig:subfig:c1}}
	\includegraphics[width=0.4\linewidth]{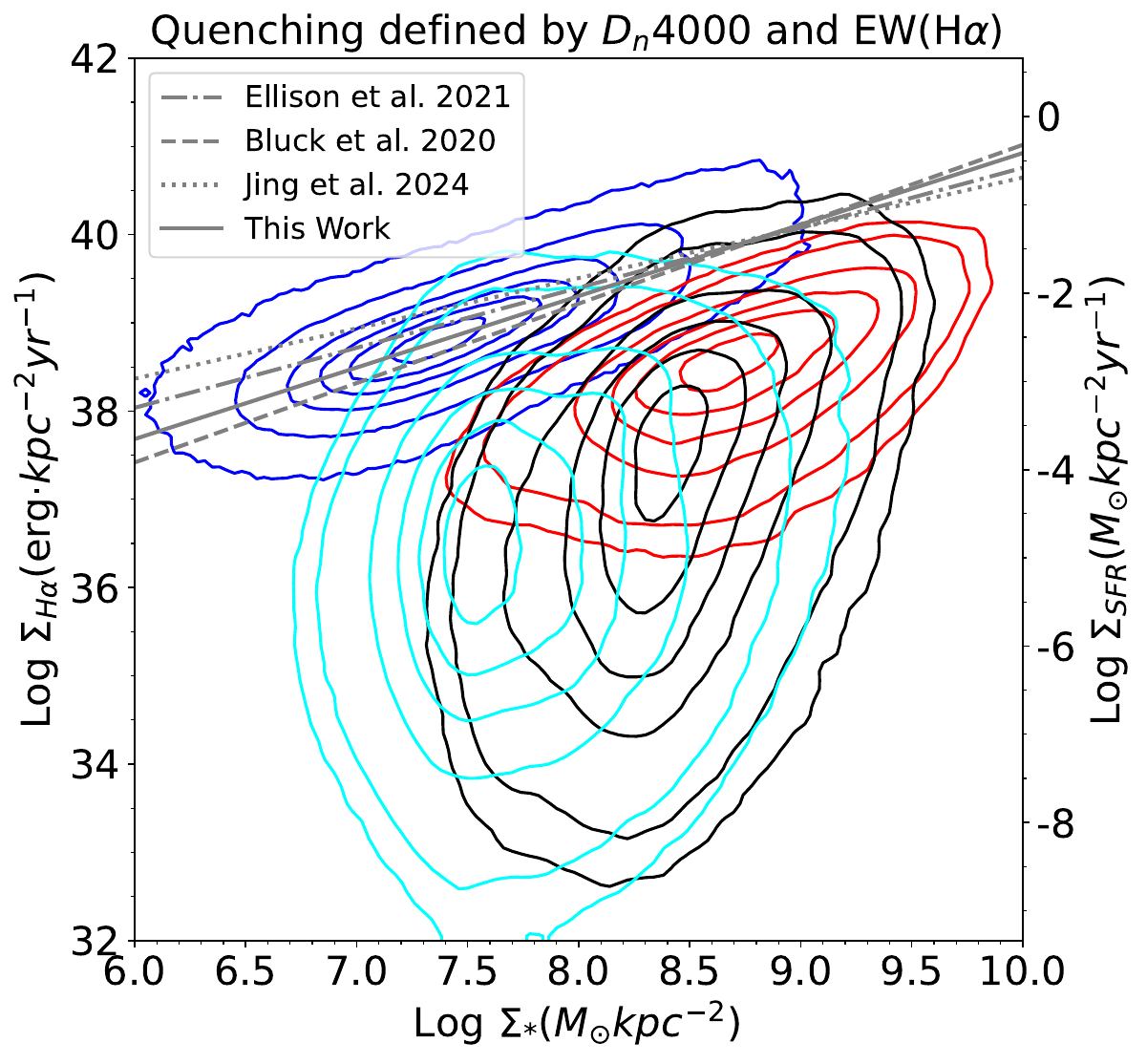}
	\hspace{0.01\linewidth}
	\subfigure{\label{fig:subfig:c2}}
	\includegraphics[width=0.4\linewidth]{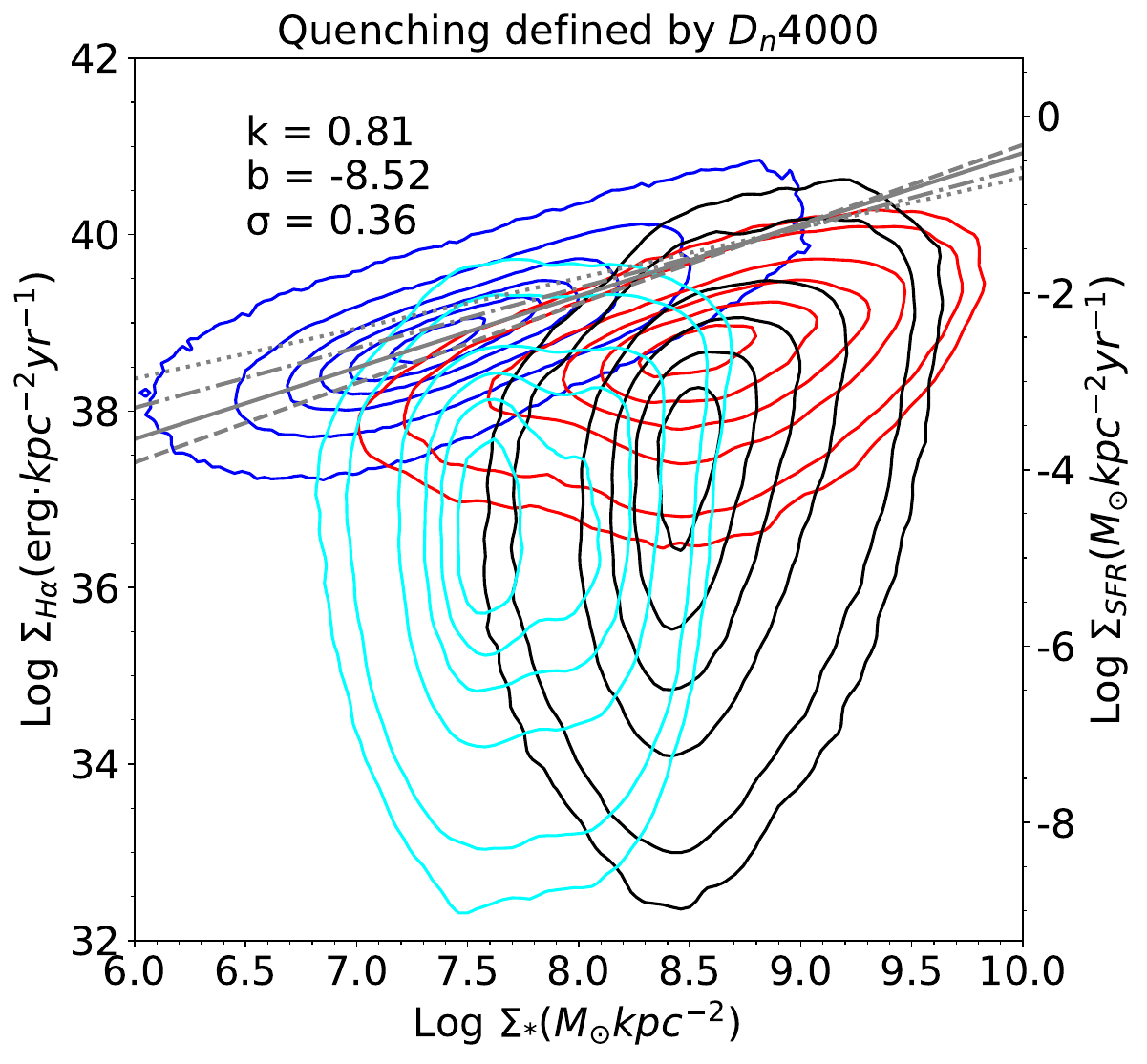}
	
\caption{Distribution of the spaxel subsamples in the plane of $\log \Sigma_{*}$ versus $\log \Sigma_{\rm H\alpha}$, with the corresponding $\log \Sigma_{\rm SFR}$ scale shown on the right axis. The quenching definitions and color coding are the same as in \autoref{fig:d4000_ha_plane}. The gray solid line in each panel represents the best-fitting rSFMS relation for the SF regions in our sample; the best-fitting slope $k$, intercept $b$, and scatter are also shown. The gray dash-dotted, dashed, and dotted comparison lines, labelled in the left panel, show the rSFMS relations of \citet{Ellison2021}, \citet{Bluck2020a}, and \citet{Jing2024}, respectively, converted to a \citet{Chabrier2003} initial mass function.}
	\label{fig:SFM_plane}
\end{figure*}

\section{Spaxel Classification} \label{sec:Q_sel}
As discussed in \autoref{sec:intro}, both the quenching definition and the data-quality requirements affect resolved sample selection. Because quenched regions are naturally weak in emission and often fail emission-line S/N cuts, we separate spaxels into star-forming(SF) regions, reliably quenched regions, and potentially quenched regions whose nominal classification is sensitive to measurement uncertainty.
\autoref{fig:sample_flowchart} summarizes the classification flow and the sample naming convention used throughout this section. We create two samples of quenched regions, with one following the scheme of Paper~II, named Sample I, and the other following the approach of \citet{Bluck2020a, Bluck2020b}, named Sample II. Under each quenching definition, we define two subsamples based on the S/N ratio, namely {\em reliably quenched} and {\em potentially quenched} regions. The classification of spaxels are not meant to be exhaustive. Spaxels which may fall in the middle between SF and quenched regions are intentionally left out for the purity of classification.
\subsection{Star-forming regions} \label{subsec:sf_regions}
We select SF regions as spaxels with \ewha $>6$\,\AA\ that fall below the \citet{Kauffmann2003} line in the BPT diagram. To construct the BPT parent sample, we require S/N $>3$ in H$\alpha$, H$\beta$, [O~{\sc iii}] $\lambda\lambda$4959,5007, and [N~{\sc ii}] $\lambda\lambda$6548,6583. This \textbf{High-SNR Parent Sample (HSPS)} contains 4,060,298 spaxels ($\sim46$\% of all spaxels), of which 3,106,523 satisfy the SF criteria.
\subsection{Reliably quenched regions} \label{subsec:q_regions}
Candidate quenched (\textbf{Q}) regions include non-SF spaxels in HSPS and spaxels in the \textbf{Low-SNR Parent Sample (LSPS)}, which cannot be classified on the BPT diagram. Because these low-S/N spaxels have larger uncertainties in \dbalmer\ and \ewha, particularly for \dbalmer, we classify a spaxel as reliably quenched only when the relevant criterion remains satisfied after accounting for measurement errors:

\begin{itemize}
 \item {\em Sample I-a:} A spaxel from HSPS is classified as quenched if it is not in the SF sample and satisfies the one-parameter criterion from Paper~II: $\dbalmer - \log(\ewha) > 1.6 - \log(2) = 1.3$. This definition is rough equivalent to the combined criteria $\dbalmer > 1.6$ and $\ewha < 2$\,\AA\ adopted in Paper~I and other studies \citep[e.g.,][]{Geha2012, Yan2012, Li2015}.
 \item {\em Sample I-b:} For spaxels from LSPS, a more conservative strategy is applied to ensure the purity of the quenched sample: $\dbalmer_{\rm low} - \log(\ewha_{\rm up}) > 1.6 - \log(2) = 1.3$. Here $\dbalmer_{\rm low}$ and $\ewha_{\rm up}$ denote the 3-$\sigma$ lower and upper limits of the $\dbalmer$ and \ewha\ measurements, respectively.
 \item {\em Sample II-a:} Following the approach of \citet{Bluck2020a,Bluck2020b}, we select regions from HSPS with low specific star formation rates (${\rm sSFR} \equiv {\rm SFR}/M_\ast$) as the quenched sample. In practice, this is implemented by requiring $D(4000) > 1.45$, based on the calibration shown in Figure~4 of \citet{Bluck2020a}. Because \citet{Bluck2020a} used the broad-definition 4000~\AA\ break whereas this series uses the narrow-definition index \dbalmer, we convert $D(4000)$ to \dbalmer\ using the scaling factor of 1.1619 proposed by \citet{Gorgas1999}. This yields an approximate threshold of $\dbalmer > 1.68$ for selecting quenched regions. This numerical threshold differs from the $\dbalmer>1.6$ criterion in Sample~I because the two selections are based on different 4000~\AA\ index definitions and calibrations.
 \item {\em Sample II-b:} Similar to Sample II-a, but with a stricter criterion applied to LSPS: $\dbalmer_{\rm low} > 1.68$.
\end{itemize}

\begin{figure*}[!thbp]
    \centering
    \subfigure{\label{fig:subfig:d1}}
	\includegraphics[width=0.48\linewidth]{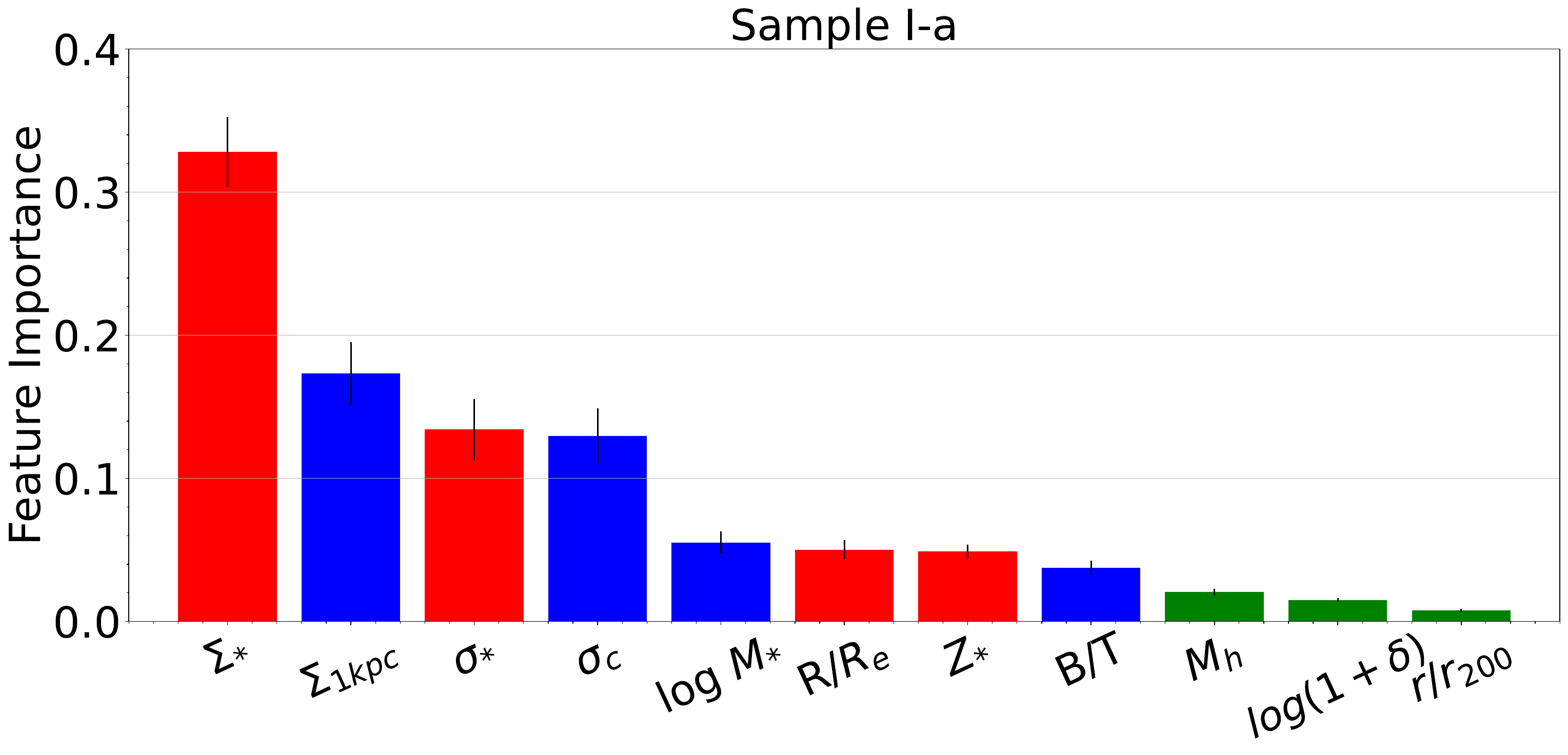}
	\hspace{0.01\linewidth}
	\subfigure{\label{fig:subfig:d2}}
	\includegraphics[width=0.48\linewidth]{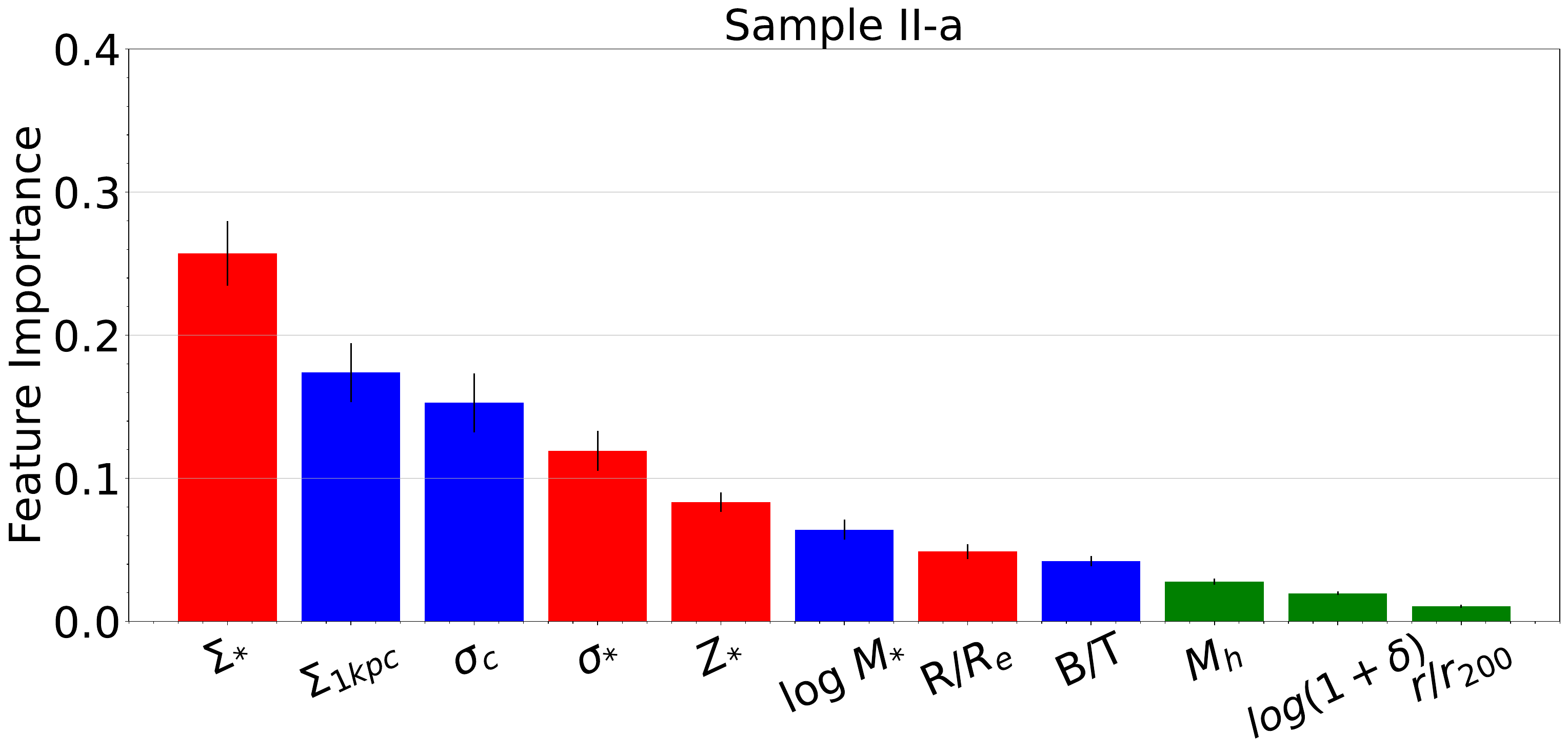}
    \caption{Gini feature importance for the HSPS-only reliably quenched samples, Sample~I-a and Sample~II-a. Bars are color-coded by parameter type: regional (red), global (blue), and environmental (green). Error bars show 1$\sigma$ uncertainties estimated by repeating the random sampling of matched SF spaxels and random forest training 100 times.}
	\label{fig:fi_hq}
\end{figure*}

\begin{figure*}[!thbp]
    \centering
    \subfigure{\label{fig:subfig:e1}}
	\includegraphics[width=0.48\linewidth]{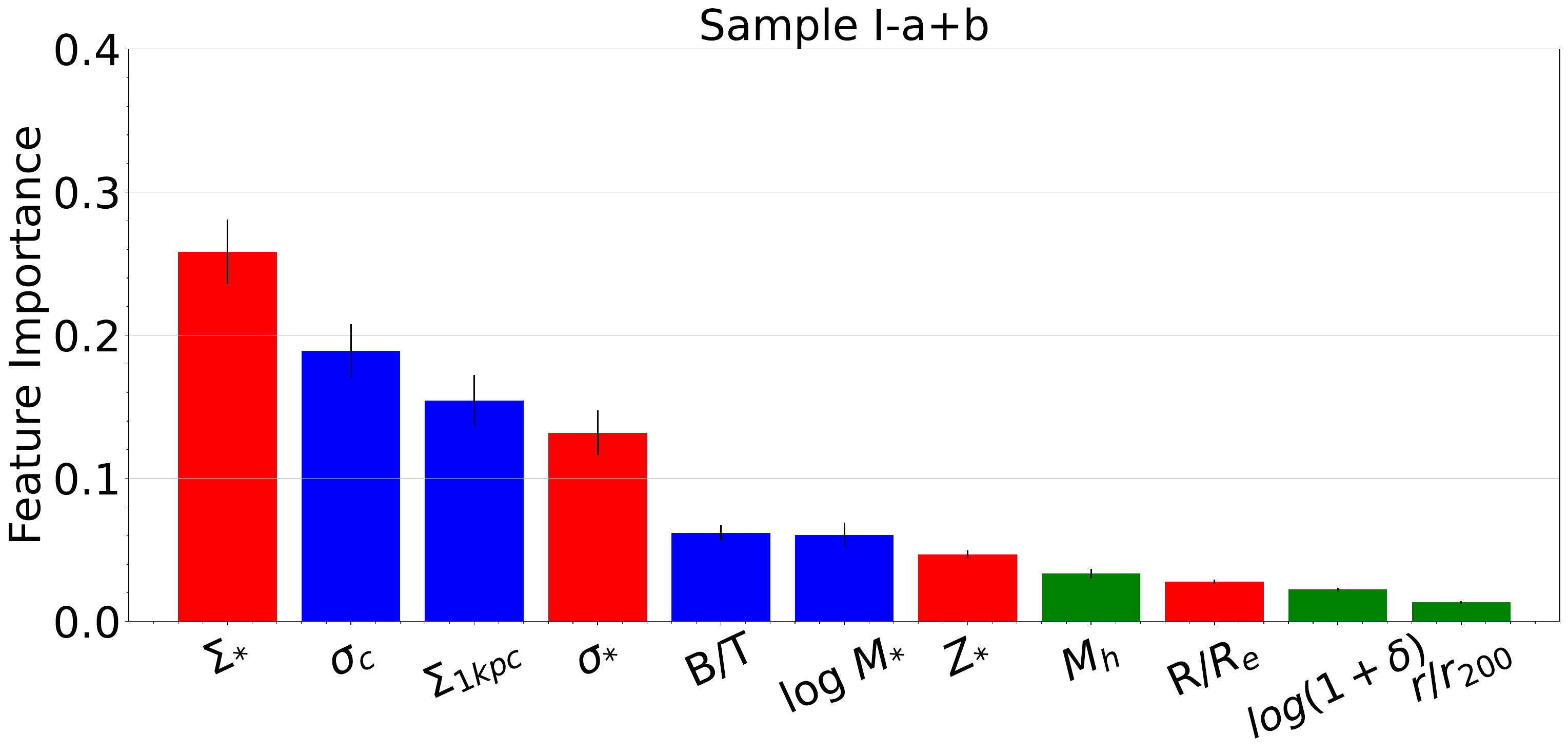}
	\hspace{0.01\linewidth}
	\subfigure{\label{fig:subfig:e2}}
	\includegraphics[width=0.48\linewidth]{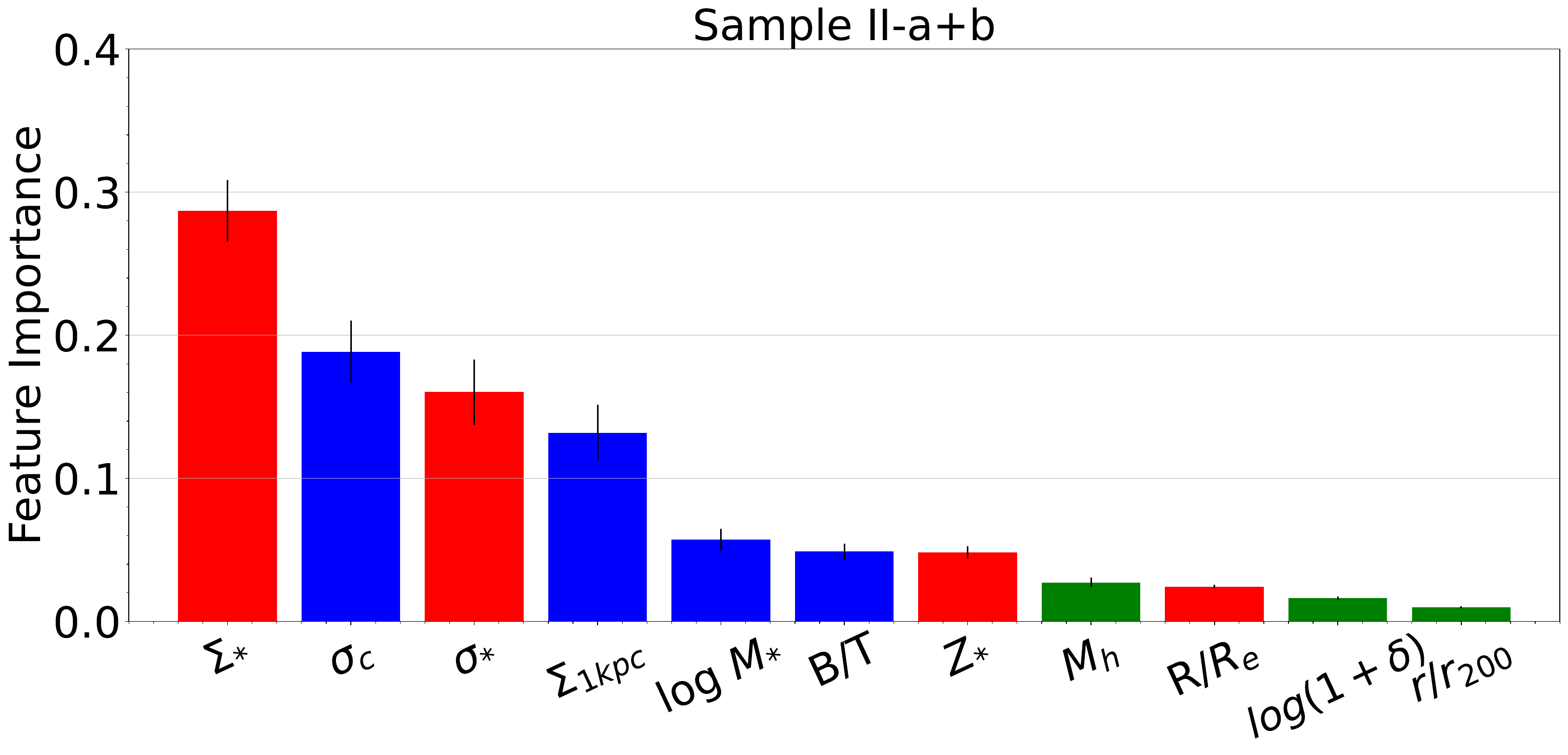}
    \caption{Gini feature importance for the reliably quenched samples selected with the two definitions, including both HSPS and LSPS spaxels (Sample~I-a+b and Sample~II-a+b). Bars are color-coded by parameter type: regional (red), global (blue), and environmental (green). Error bars show 1$\sigma$ uncertainties estimated by repeating the random sampling of matched SF spaxels and random forest training 100 times.}
	\label{fig:fi_reliable}
\end{figure*}

\subsection{Potentially quenched regions} \label{subsec:p_regions}
After identifying SF and reliably quenched regions, $\sim$4 million LSPS spaxels remain unclassified. Some satisfy the nominal quenching criteria using best-fit values but fail the error-aware requirements above. We call these {\em potentially quenched regions}; they indicate how analyses without the same uncertainty treatment may mix reliable and ambiguous regions:
\begin{itemize}
 \item {\em Sample I-c:} LSPS spaxels that cannot be classified as either SF or reliably quenched, but satisfy $\dbalmer - \log(\ewha) > 1.6 - \log(2) = 1.3$ using the best-fit measurements.
 \item {\em Sample II-c:} Similar to Sample I-c, LSPS spaxels that cannot be classified as either SF or reliably quenched, but satisfy $\dbalmer > 1.68$ using the best-fit measurements. We note that \citet{Bluck2020a} did not account for measurement errors in \dbalmer; consequently, the analogue of Sample~II-c would have been included in their quenched sample.
\end{itemize}

\autoref{fig:d4000_ha_plane} shows all subsamples selected above [Sample I-a(b,c) and II-a(b,c), from left to right] on the diagram of $\log_{10}[{\rm EW}({\rm H}\alpha)]$ versus \dbalmer. For comparison, the parent sample of all spaxels is shown as gray contours in the background, with star-forming regions as blue contours. The two definitions select somewhat different subsets within HSPS (red contours): Sample II-a contains more spaxels with relatively higher \ewha, with approximately 33\% exceeding the demarcation line of Sample I. For quenched regions in LSPS (black contours), both definitions yield similar results and occupy the high-\dbalmer, low-\ewha region. The potentially quenched samples (cyan contours) contain much more spaxels than the reliably quenched samples and partially overlap with the reliably quenched regions. 

In \autoref{fig:SFM_plane}, we show the same subsamples on the $\Sigma_{\rm H\alpha}$--$\Sigma_\ast$ plane. The right-hand axis gives the corresponding $\Sigma_{\rm SFR}$ based on the \citet{Kennicutt1998} calibration, and the gray line shows the best-fit rSFMS for our SF sample. The overplotted literature rSFMS relations provide a reference for the normalization and slope of our SF sequence; small offsets are expected because of differences in sample selection, SFR calibration, and fitting method. Reliably quenched regions lie distinctly below the rSFMS, with LSPS quenched regions showing especially low $\Sigma_{\rm H\alpha}$. Potentially quenched regions occupy a broader distribution and overlap substantially with the rSFMS, motivating their separate treatment in the feature-importance analysis.

\section{Results} \label{sec:results}

We use a Random Forest classifier, implemented in \textsc{Scikit-learn} \citep{Breiman2001, Pedregosa2011}, to distinguish quenched from SF regions. Following Paper~II, we use Gini feature importance as a measure of predictive association rather than direct causality. All random forest analyses use the 11 input properties defined in \autoref{subsec:data_pre} and \autoref{subsec:host_env}. These properties are chosen to closely match \citet{Bluck2020a}, so that differences in feature importance can be compared primarily in terms of quenching definition and sample selection. We do not include stellar age, \dbalmer, \ewha, or N2H$\alpha$ in the classifier because some enter the sample definitions directly, while Paper~II showed that N2H$\alpha$ and stellar age are better interpreted as tracers or outcomes rather than independent candidate drivers.

To mitigate class imbalance, we randomly draw a matched SF sample equal in size to each quenched sample. The combined sample is used to train the Random Forest classifier, with 20\% held out as a test set. We repeat the full procedure 100 times to estimate uncertainties in the feature importance scores.

\subsection{Feature importance for reliably quenched regions} \label{subsec:fi_h}

\begin{figure*}[t!]
    \centering
    \subfigure{\label{fig:subfig:f1}}
	\includegraphics[width=0.48\linewidth]{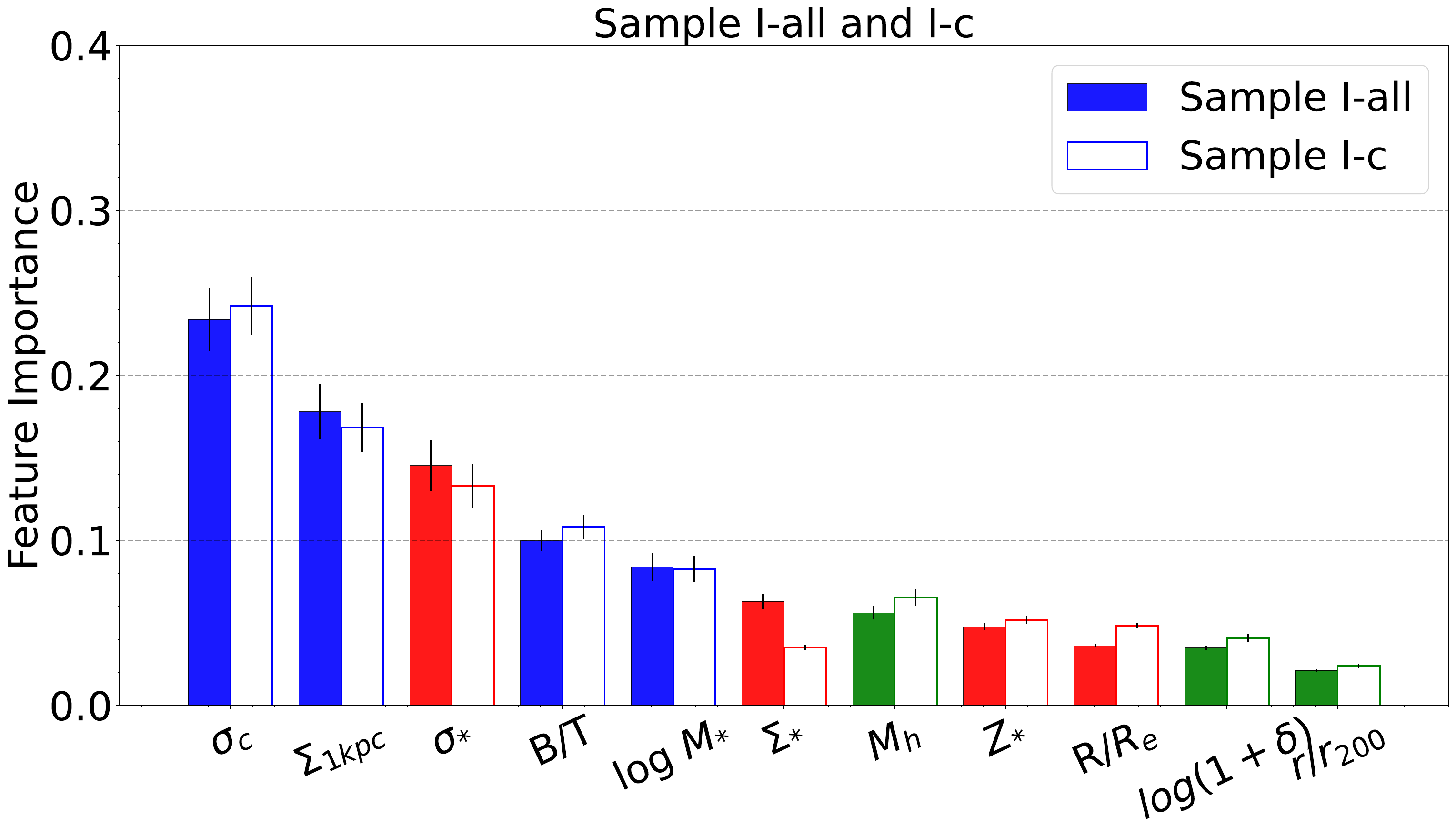}
	\hspace{0.01\linewidth}
	\subfigure{\label{fig:subfig:f2}}
	\includegraphics[width=0.48\linewidth]{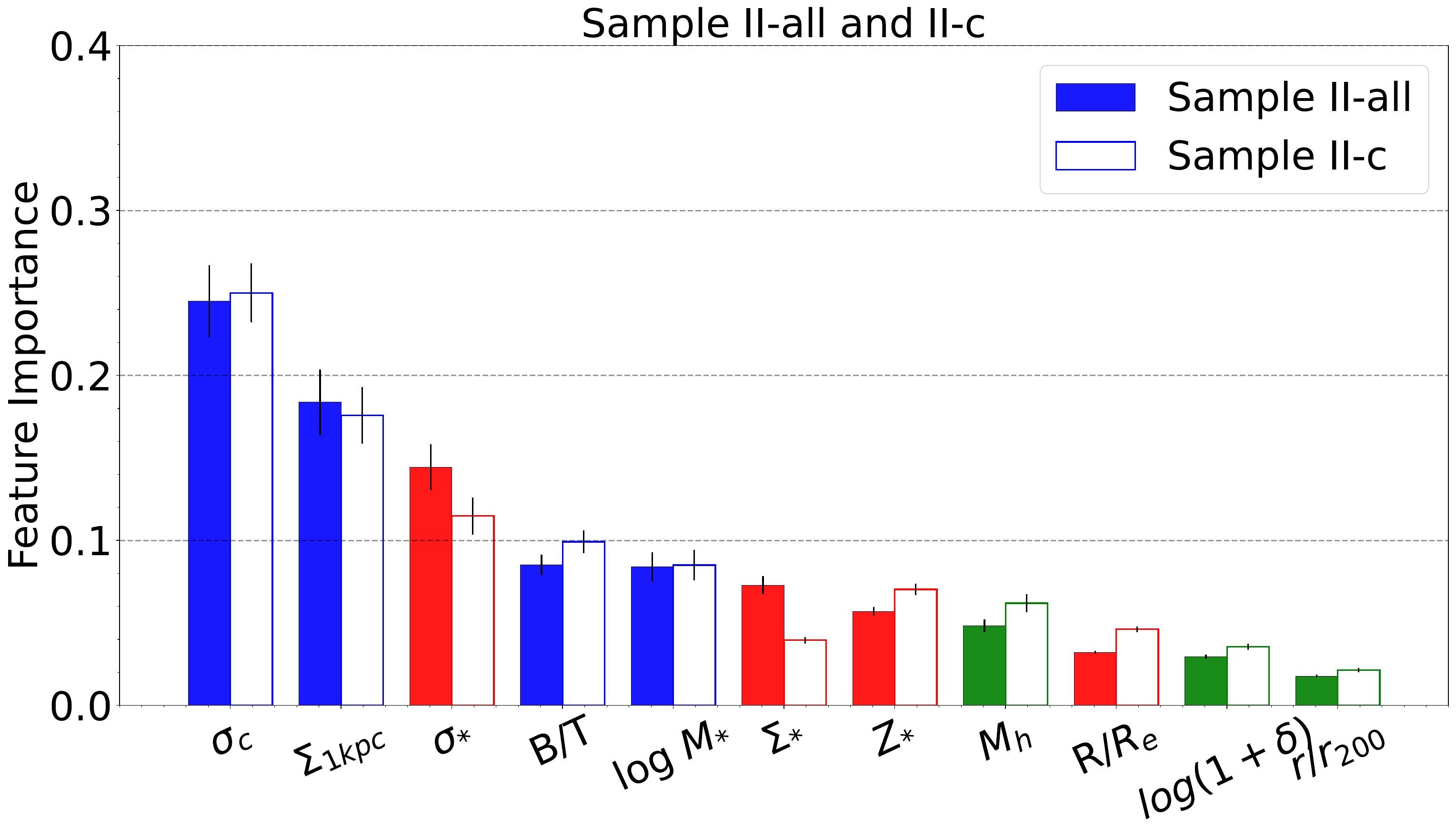}
    \caption{Gini feature importance after including potentially quenched regions. Solid bars show the results when reliably quenched and potentially quenched regions are combined, while hollow bars show the potentially quenched samples alone. Colors and error bars are the same as in \autoref{fig:fi_reliable}.}
	\label{fig:fi_potential}
\end{figure*}

We first consider Sample~I-a, which follows the Paper~I/Paper~II quenching definition and uses only HSPS spaxels with reliable emission-line classifications. The left panel of \autoref{fig:fi_hq} shows that $\Sigma_\ast$ is the most important feature, consistent with the main result of Paper~II. This agreement is an important baseline: using a larger and more general MaNGA sample, the same local quantity remains most strongly associated with reliably quenched regions. We then add the error-vetted low-S/N spaxels in Sample~I-b. The left panel of \autoref{fig:fi_reliable} shows that $\Sigma_\ast$ remains dominant for Sample~I-a+b, indicating that including reliably quenched LSPS regions does not change the result. This test was not examined in Paper~II, where the analysis was limited to a cleaner disk-galaxy sample.

\begin{figure*}[ht!]
    \centering
    \includegraphics[width=\linewidth]{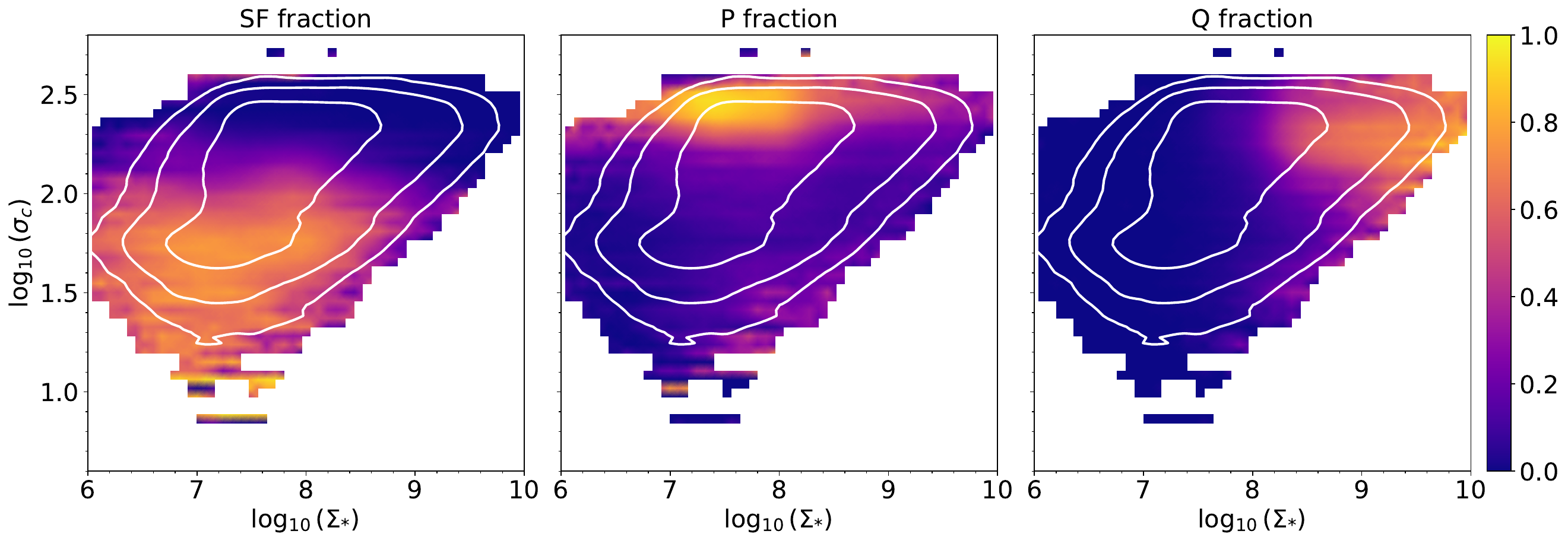}
    \caption{The $\Sigma_\ast$ versus $\sigma_c$ diagram color-coded by the fraction of SF, potentially quenched (P), and reliably quenched (Q) spaxels in each bin. Contours enclose 68\%, 95\%, and 99\% of the data.}
    \label{fig:sigm_sigc}
\end{figure*}

It is more surprising that the same conclusion is obtained with the Bluck-like quenching definition. For the HSPS-only Sample~II-a, $\Sigma_\ast$ is again the leading feature (right panel of \autoref{fig:fi_hq}). After adding the error-vetted LSPS spaxels, the right panel of \autoref{fig:fi_reliable} shows that $\Sigma_\ast$ still has substantially higher importance than any other parameter. Because Sample~II was designed to approximate the \citet{Bluck2020a,Bluck2020b} selection, one might have expected global quantities such as $\sigma_c$ or $\Sigma_{\rm 1kpc}$ to become dominant. Instead, for reliably quenched regions, the dominance of $\Sigma_\ast$ does not depend on either the quenching definition or the inclusion of low-S/N spaxels.

\begin{figure*}[t!]
    \centering
    \subfigure{\label{fig:subfig:g1}}
	\includegraphics[width=0.48\linewidth]{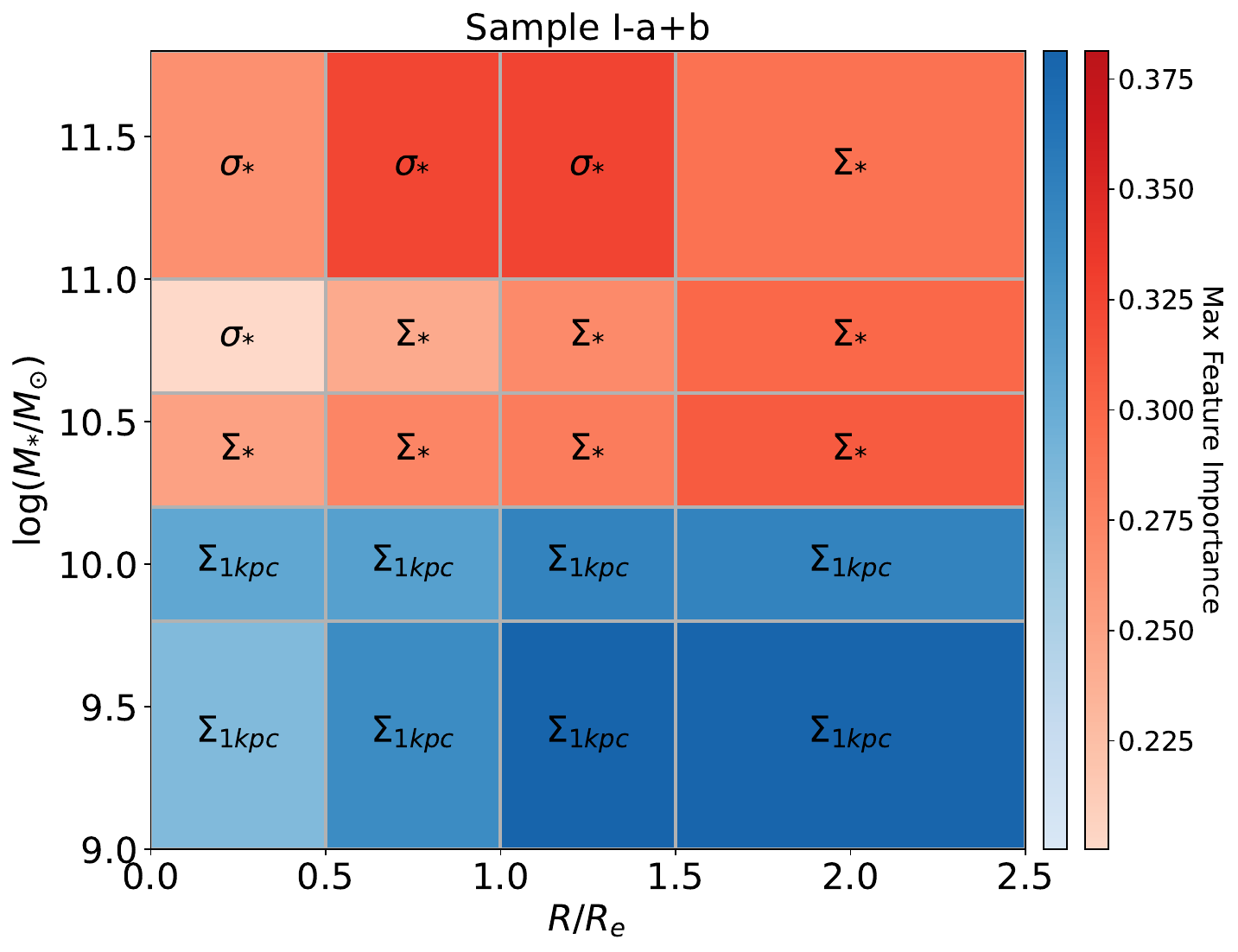}
	\hspace{0.01\linewidth}
	\subfigure{\label{fig:subfig:g2}}
	\includegraphics[width=0.48\linewidth]{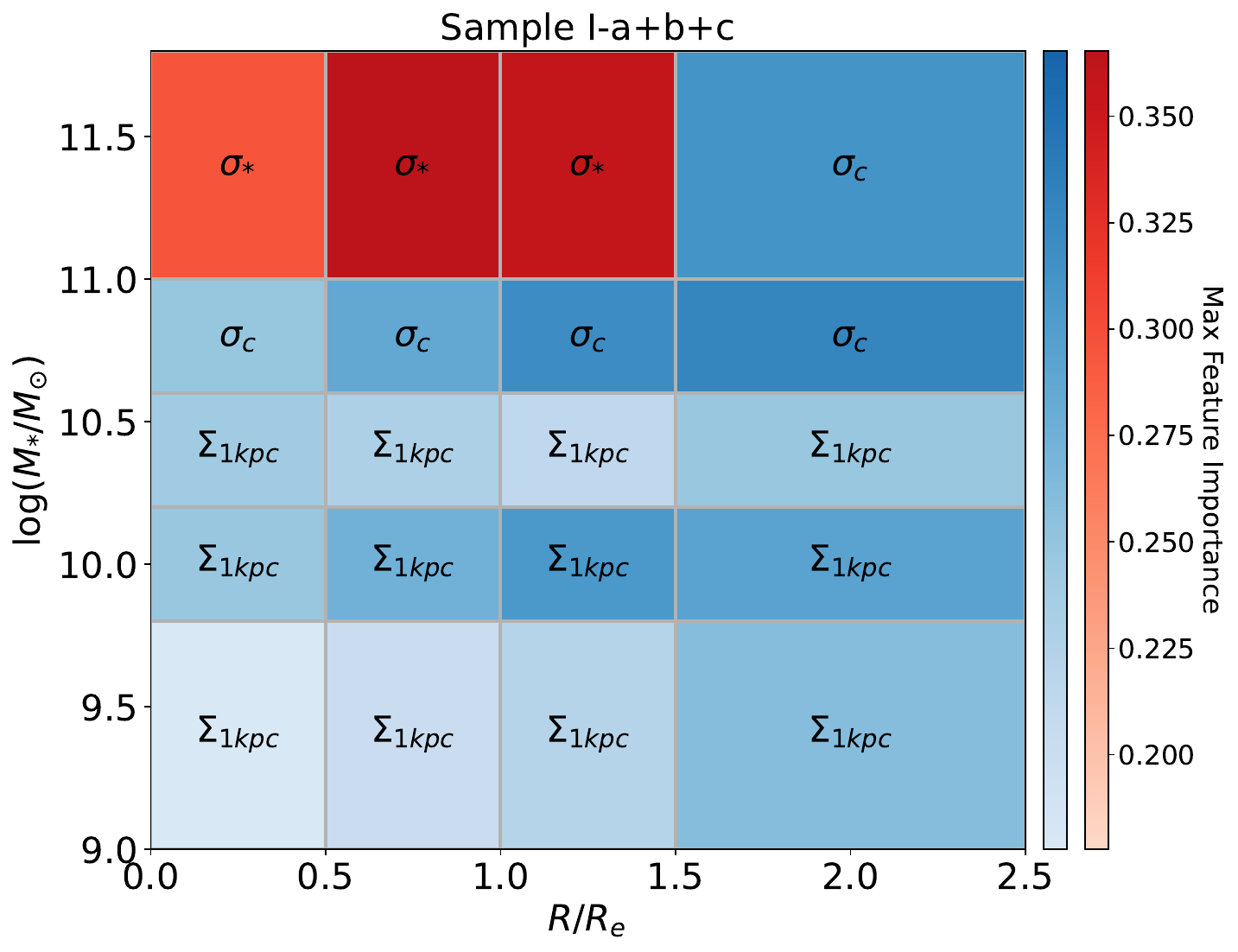}
    \caption{Heatmap of the maximum feature importance in different global stellar mass and galactocentric radius bins. In each cell, the name of the dominant parameter is indicated. The color bars represent the feature importance values, with red and blue denoting regional and global properties, respectively. The left panel shows the results for reliably quenched regions only (Sample I-a and I-b), while the right panel additionally includes potentially quenched regions (Sample I-c).}
	\label{fig:fi_grid}
\end{figure*}

\subsection{The role of potentially quenched regions} \label{subsec:fi_p}
The main difference between our results and those of \citet{Bluck2020a} is the treatment of potentially quenched regions. As noted in \autoref{subsec:p_regions}, our Sample~II-c would enter the quenched sample if measurement uncertainty in \dbalmer\ were not considered. \autoref{fig:fi_potential} shows that $\sigma_c$ and $\Sigma_{\rm 1kpc}$ dominate for the potential samples alone, and that combining potential and reliable regions recovers the \citet{Bluck2020a} result that $\sigma_c$ has the highest feature importance. Thus the high importance of $\sigma_c$ is tied mainly to the potentially quenched population, regardless of the quenching definition.

The reason is visible in \autoref{fig:SFM_plane}: potentially quenched regions overlap much more strongly with the rSFMS than reliably quenched regions do, so mixing the two changes the composition of the quenched sample. The much larger sample size of potentially quenched regions also overwhelms the reliably quenched sample and changes the feature importance order. \autoref{fig:sigm_sigc} shows the same point in the $\Sigma_\ast$--$\sigma_c$ plane for Sample~I. Potentially quenched regions are concentrated at high $\sigma_c$ but have $\Sigma_\ast$ values similar to SF regions, with most below $10^8\,\msun\,{\rm kpc}^{-2}$. Reliably quenched regions instead appear mainly above $\Sigma_\ast \gtrsim 10^8\,\msun\,{\rm kpc}^{-2}$, and their fraction approaches unity above $10^9\,\msun\,{\rm kpc}^{-2}$. This contrast explains why $\sigma_c$ dominates in the potential samples whereas $\Sigma_\ast$ dominates for reliable quenched regions. We also emphasize that these potentially quenched regions could well be contaminated significantly by SF regions or intermediate regions, as they can be scattered there due to low S/N in the measurements. Considering the large number of SF spaxels and potential samples, the $\sigma_c$ may be related to the transition between SF and non-SF ionization mechanisms, rather than being related to quenching.

Feature-importance rankings based on mixed samples should therefore not be interpreted as evidence for a single global quenching pathway. The high importance of $\sigma_c$ and $\Sigma_{\rm 1kpc}$ indicates that they play an important role in the cessation process, but may not be directly responsible for the production of the reliably quenched population.

\subsection{Dependence on stellar mass and galactocentric radius} \label{subsec:massrole}

The leading parameter may also depend on stellar mass and location within a galaxy. Paper~I identified a critical mass near $\sim10^{10.2}\,\msun$: above this scale, galaxies tend to quench from the inside out, while lower-mass galaxies evolve more synchronously across radius \citep{Wang2018}.

We therefore divide the SF and quenched samples into 20 bins: five stellar-mass intervals ($\lgmstar<9.8$, $9.8<\lgmstar<10.2$, $10.2<\lgmstar<10.6$, $10.6<\lgmstar<11.0$, and $\lgmstar>11.0$) and four radial intervals ($R/R_e<0.5$, $0.5<R/R_e<1.0$, $1.0<R/R_e<1.5$, and $1.5<R/R_e<2.5$). We show only the results for Sample~I here, as Sample~II yields similar conclusions. As shown in the left panel of \autoref{fig:fi_grid} which is based on the reliably quenched regions, the dominant parameter changes strongly with mass: $\Sigma_{\rm 1kpc}$ leads at $\lgmstar<10.2$ across all radii, whereas local properties $\Sigma_\ast$ and $\sigma_\ast$ become more important at high mass. In the outer regions of massive galaxies, $\Sigma_\ast$ remains the leading parameter, showing that purely global interpretations are incomplete for reliably quenched regions. Overall, this recovers the Paper~I transition near $\sim10^{10.2}\,\msun$: local properties become prominent above this mass, while $\Sigma_{\rm 1kpc}$ is more important below it. 

When potentially quenched regions are additionally considered, the right panel of \autoref{fig:fi_grid} shows that the global properties, $\sigma_c$ or $\Sigma_{\rm 1kpc}$ take the place of the leading parameters in all cases except the inner regions ($R<1.5R_e$) of the highest mass galaxies ($\lgmstar>11$), for which the local property $\sigma_\ast$ is still the most important. This result reinforces the dominating role of the potential population in studies of quenching.

\section{Discussion} \label{sec:discussion}
\subsection{Possible interpretations of \texorpdfstring{$\Sigma_{\rm 1kpc}$}{Sigma 1kpc}} \label{subsec:sig1}

$\Sigma_{\rm 1kpc}$ traces the central stellar mass surface density and is known to correlate with galaxy color, quenching thresholds, bulge structure, black-hole mass proxies, and cold-gas depletion \citep{Cheung2012, Fang2013, Chen2020, Cattaneo2025, Shread2026}. These connections often motivate an interpretation in terms of bulge growth, AGN feedback, or morphological quenching.

It is therefore notable that $\Sigma_{\rm 1kpc}$ is most important at low mass in our analysis. If it is read as an AGN-related global property, the result would imply that AGN feedback may matter in low-mass galaxies, where it is not usually expected to dominate. Recent observations and simulations do suggest that AGN-driven outflows can affect dwarf or low-mass systems \citep[e.g.][]{Smethurst2016, Penny2018, Baldassare2020, Silk2017, Dashyan2018, Arjona2024, Salehirad2025}, but the role of such feedback remains uncertain.

Another possibility is that high $\Sigma_{\rm 1kpc}$ is partly an outcome of quenching. Previous work found that satellite and low-mass galaxies can move to higher $\Sigma_{\rm 1kpc}$ during quenching, for example if star formation is suppressed first in the outskirts while central star formation is enhanced due to galaxy–galaxy interactions or external ram pressure\citep{Woo2017, Guo2021}. This interpretation is plausible because low-mass galaxies are more vulnerable to environmental effects, even though the environmental parameters used here have low feature importance. A high $\Sigma_{\rm 1kpc}$ may encode the outcome of environmental or morphological processes rather than the external driver itself. Spatially resolved gas observations will be needed to distinguish these possibilities.

\subsection{Dominance of \texorpdfstring{$\sigma_c$}{sigma c} in potentially quenched regions} \label{subsec:sigc}

The high importance of $\sigma_c$ in previous work is associated mainly with the potentially quenched regions defined here. These spaxels have lower H$\alpha$ surface brightness than SF regions but similar $\Sigma_\ast$, unlike reliably quenched regions. Although they occupy similar positions on the \dbalmer--\ewha\ diagnostic diagram, their random-forest behavior points to different associated properties. A connection with AGN-related ionization or feedback is possible, but not unique.

Caution is especially important because Paper~II showed that regions selected without a low-\ewha\ requirement can include genuinely quenched regions with AGN ionization, mixed-ionization regions, or star-forming regions whose young populations are difficult to recover from optical spectra alone. The potential samples here may similarly trace an earlier or more ambiguous phase rather than the same physical state as reliably quenched regions. Higher-resolution observations and ionization-source decomposition will be required to separate these cases \citep[e.g.][]{Teimoorinia2024}.

\subsection{High importance of \texorpdfstring{$\Sigma_\ast$}{Sigma star}} \label{subsec:Sigma_m}

The high importance of $\Sigma_\ast$ found in Paper~II was based on a limited disk-galaxy sample. Here we show that the result extends to a more general population: for reliably quenched regions, especially in massive galaxies, $\Sigma_\ast$ is the dominant local parameter regardless of quenching definition.

High $\Sigma_\ast$ may help maintain quiescence through several local processes involving evolved stars \citep[e.g.][]{Toomre1964, Wang1994, Kormendy2004, Tasker2009, Raskutti2016, Kim2017, Pathak2025}. Stellar winds and supernovae can affect the local fuel supply, while the stellar gravitational potential, shear, and radiation pressure can inhibit cloud collapse. Recent observations combining warm ionized gas and molecular gas support a role for shear and radiation pressure in suppressing star formation in early-type galaxies \citep{Lu2025}, consistent with the Paper~II emphasis on evolved-star radiation pressure \citep{Jing2024}. At the same time, $\Sigma_\ast$ is correlated with star formation activity in ways that can depend on sample, environment, and scale \citep[e.g.][]{Shi2011, Shi2018, Lin2019a, Pessa2022, Ellison2024}. Gas observations will therefore be needed to determine when high $\Sigma_\ast$ is a cause, a consequence, or a maintainer of the quenched state.

\section{Summary}\label{sec:sum}

This study builds on Papers~I and II by testing how the definition of quenched regions affects feature-importance results in MaNGA galaxies. We classify spaxels as star-forming, reliably quenched, or potentially quenched while accounting for measurement uncertainties, then train random forest classifiers using a parameter set chosen for direct comparison with \citet{Bluck2020a}. We compare two quenching definitions and examine how the results vary with stellar mass and galactocentric radius.

Our conclusions can be summarized as follows:
\begin{itemize}
 \item Based on reliably quenched regions, the high importance of $\Sigma_\ast$ found in Paper~II extends to a more general galaxy population beyond the disk component of isolated late-type galaxies.

 \item The discrepancy with \citet{Bluck2020a} is largely associated with potentially quenched regions. The high importance of $\sigma_c$ is recovered when these ambiguous regions are included. For reliably quenched regions, both quenching definitions instead identify $\Sigma_\ast$ as the most important parameter.
  
 \item For reliably quenched regions, the parameters most closely associated with the quenched state depend strongly on global stellar mass. Above the critical stellar mass of $\sim10^{10.2}\,\msun$, local properties such as $\Sigma_\ast$ and $\sigma_\ast$ become prominent, while the global property traced by $\Sigma_{\rm 1kpc}$ is more important at lower masses. 
\end{itemize}

Overall, quenching criteria and measurement uncertainties strongly affect the interpretation of feature-importance analyses. The dominance of $\sigma_c$ in samples including potentially quenched regions may point to AGN-related or mixed-ionization phenomena, but not to a single universal quenching mechanism. Reliably quenched regions are closely linked to high local stellar mass surface density ($\Sigma_\ast \gtrsim 10^8\,\msun\,{\rm kpc}^{-2}$), especially in massive galaxies. Spatially resolved gas observations and ionization-source decomposition will be needed to identify the underlying mechanisms.
\begin{acknowledgments}
This work is supported by the National Science Foundation of China (grant No. 12425302, 12373008, 12433003) and the National Key R\&D Program of China (grant No. 2018YFA0404502).

Funding for SDSS-IV has been provided by the Alfred P. Sloan Foundation and Participating Institutions. Additional funding for SDSS-IV has been provided by the US Department of Energy Office of Science. SDSS-IV acknowledges support and resources from the Center for High-Performance Computing at the University of Utah. The SDSS web site is www.sdss.org.

We acknowledge the Tsinghua Astrophysics High-Performance Computing platform at Tsinghua University for providing computational and data storage resources that have contributed to the research results reported within this paper.
\end{acknowledgments}

\software{Scikit-learn \citep{Pedregosa2011}}

\appendix

\section{Appendix: $\Sigma_{\rm 1kpc}$ Comparison} \label{sec:app}
\begin{figure}[ht]
    \centering
    \includegraphics[width=\linewidth]{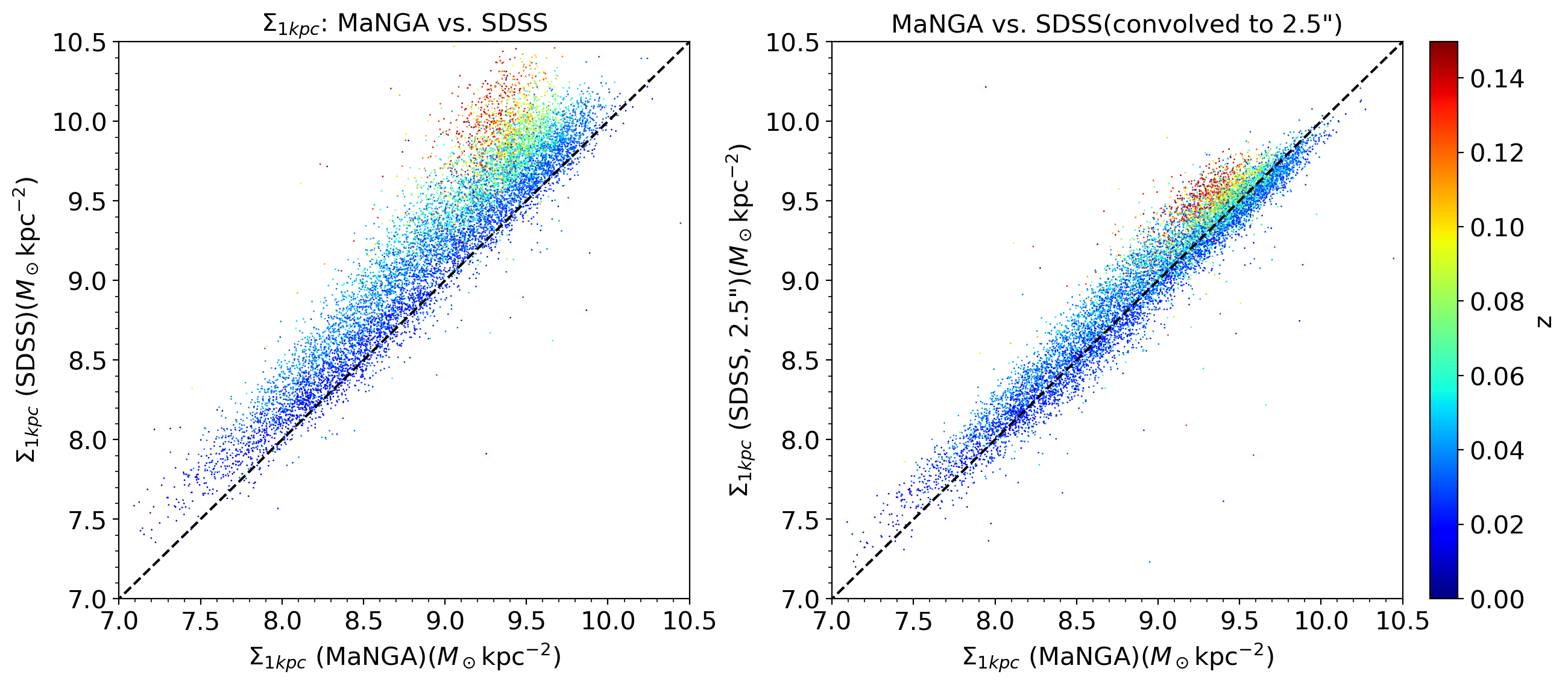}
    \caption{Comparison of $\Sigma_{\rm 1kpc}$ derived from MaNGA data versus SDSS imaging-based estimates, color-coded by the galaxy redshift. The left panel corresponds to the original SDSS images, with a typical spatial resolution of $\sim1.4^{\prime\prime}$; the right panel displays the results after convolving the images to a $2.5^{\prime\prime}$ resolution, matching that of the MaNGA observations.}
    \label{fig:sigma1_com}
\end{figure}

In the main text we use $\Sigma_{\rm 1kpc}$ derived from SDSS imaging rather than that from the MaNGA data. This is because the relatively poor spatial resolution of MaNGA ($\sim2.5^{\prime\prime}$) leads to significant underestimation of $\Sigma_{\rm 1kpc}$ for galaxies at $z\ga 0.03$, as shown in the left panel of~\autoref{fig:sigma1_com}. The right panel demonstrates the consistency between the two measurements after degrading the SDSS imaging resolution from $\sim1.4^{\prime\prime}$ to the MaNGA resolution.

\bibliography{sample7}{}
\bibliographystyle{aasjournalv7}



\end{document}